"\pdfoutput=1"
\documentclass[12pt]{article}

 \usepackage{xcolor}
\usepackage[centertags]{amsmath}
\usepackage{amsthm}
\usepackage{amsfonts}
\usepackage{amssymb}
\usepackage{newlfont}
\usepackage{delarray}
\usepackage{graphicx}
\usepackage[numbers,sort&compress]{natbib}

\begin{document}
  \title{Two-stage filtering for joint state-parameter estimation}
\author{
  Santitissadeekorn, Naratip\\
  \texttt{naratip.math@gmail.com}
  \and
  Jones, Chris\\
  \texttt{ckrtj@email.unc.edu}
}

 \maketitle

\begin{abstract}
{This paper presents an approach for simultaneous estimation of the state and unknown parameters in a sequential data assimilation framework.
The state augmentation technique, in which the state vector is augmented by the model parameters, has been investigated in many previous studies and some success with this technique has been reported in the case where model parameters are additive. However, many geophysical or climate models contains non-additive parameters such as those arising from physical parametrization of sub-grid scale processes, in which case the state augmentation technique may become ineffective since its inference about parameters from partially observed states based on the cross covariance between states and parameters is inadequate if states and parameters are not linearly correlated. In this paper, we propose a two-stages filtering technique that runs particle filtering (PF) to estimate parameters while updating the state estimate using Ensemble Kalman filter (ENKF; these two ``sub-filters" interact. The applicability of the proposed method is demonstrated using the Lorenz-96 system, where the forcing is parameterized and the amplitude and phase of the forcing are to be estimated jointly with the states. The proposed method is shown to be capable of estimating these model parameters with a high accuracy as well as reducing uncertainty while the state augmentation technique fails.}
\end{abstract}
\newpage

\section{Introduction}\label{sec:intro}
 The usefulness and reliability of a data assimilation (DA) technique typically requires a mathematical model that accurately simulate the actual dynamical processes. In many instances, the model contains uncertain parameters which may appear as additive or non-additive parameters or the so-called ``closure parameters", which arise from the parameterizations of the unresolved sub-scale processes. Use of the incorrect values of the parameters in the DA may lead to large errors in the state estimates and inconsistency between the forecast and reality. A key strategy in increasing effectiveness of numerical prediction of climate, weather or other geophysical processes is the development of a DA method for simultaneously estimating values of the model parameters as well as the state variables that are both incompletely known. The problem of the joint state-parameter estimation has been investigated in many previous works. To deal with the uncertainty of model parameter in the context of DA, a commonly used DA approach such as the ensemble Kalman filter (ENKF)~\cite{Evensen94,HM98} or local ensemble transform Kalman filter (LETKF)~\cite{LETKF} has been adapted by augmenting the state vector with the uncertain parameters; hence the augmented method~\cite{BHKOS06,Gillijns2007,BBM2013}. The standard Kalman update equations are then applied to estimate the combined state-parameter vector. In all Kalman-type methods, the inference of the parameters relies substantially on the (flow-dependent) cross-covariance between the states variables and the model parameters, which is approximated from the ensemble forecast in the ensemble-based methods. If the dimension of the model parameters is comparable to that of the state vector, the augmented state vector becomes significantly larger than that of the original problem, which introduces the increase in the computational load as well as the inaccuracies in computing covariance matrices. One approach to avoid this difficulty is the interacting Kalman filter whereby two Kalman filters are designed to estimate states and parameter separately and the two filters interact~\cite{Friedland1969,MSGH2005, KW2010}. A more recent approach to this problem is the use of the augmented LETKF~\cite{BHKOS06, BBM2013}, which computes in parallel the Kalman update equations for the subdivided local regions with a smaller dimensions of the state vector. Both approaches demonstrate successful results in the case of the additive parameters. Unfortunately, in the cases of the multiplicative parameters, the augmented techniques are usually problematic as demonstrated in Yang~\cite{YangDelsole2009}.\\
\par
 In this paper, we focus on the case that the dimension of the state vector is large while that of the non-additive model parameters is comparatively small. Hence, we apply the ENKF to estimate the state and separately estimate the model parameters by the particle filtering (PF) and the two sub-filters recursively interact; hence the name ``two-stage" filtering. The results will be compared with the augmented ENKF, which will be reviewed in Section~\ref{sec:Augmented}, to confirm the ineffectiveness of using it in such situation. In a way, the two-stage filtering method in this paper can be considered as a suboptimal ``approximation" of the well-known Rao-Blackwellized particle filtering (RBPF)~\cite{RBPF99,DGA2000} and this will be explained in Section~\ref{sec:twostage}. In section~\ref{sec:case1}, we test the proposed method using the Lorenz-96 model~\cite{Lorenz96} and assume the ``perfect model" scenario where the only source of the model error is the uncertain parameters. Section~\ref{sec:case2} addresses the``imperfect model" case using the fast-slow Lorenz-96 model as a proof of concept in which the closure parameters arising from some parameterizations of the fast-scale process
 will be estimated.

\section{Augmented ENKF for joint state-parameter estimation}\label{sec:Augmented}
Let $x\in\Re^m$ be the $m-$ dimensional model state vector and $\theta\in\Re^q$ be the $q-$dimensional vector specifying the model parameter whose true values are constant but unknown. Let $f_{k}^{k+1}(x(t_k),\theta)$ be a map that propagate the state at time $t_k$ to $t_{k+1}$. In the augmented method, we treat the combined vector $w=[x,\theta]^T\in\Re^{n+q}$ as the new state vector that is updated according to a dynamical system
\begin{equation}\label{eq:model}
w_{k+1}= \tilde{f}_{k}^{k+1}(w_k)=\left[\begin{array}{c}f_{k}^{k+1}(x_k,\theta_k)\\\theta_k\end{array}\right].
\end{equation}
Let $y\in\Re^r$ be the $r-$dimensional observation vector which is related to the model state by the equation
\begin{equation}
y_{k+1}=\mathbf{H}x+\varepsilon_k,
\end{equation}
where $\varepsilon_k$ is assumed to be zero-mean Gaussian noise with covariance matrix $\mathbf{R}$ and the observation operator $\mathbf{H}\in\Re^{r\times m}$ is assumed to be linear only to simplify notation but our discussion below is still valid without this assumption. In most situations, the model parameters are not observed and the observation operator for the augmented system has a form
\begin{equation}
\tilde{\mathbf{H}}w=[\mathbf{H}\quad0]w=\mathbf{H}x.
\end{equation}

In the ENKF method, the spread of the ensemble of size $n$ is used to approximate the background error covariance matrix and the Kalman update equations are applied to approximate the analyzed ensemble mean and the analysis error covariance matrix. For the above augmented state-parameter system, the background error covariance has the following sub-structures.
\begin{equation}\label{eq:Bk}
\mathbf{P}=\begin{bmatrix}\mathbf{P}_{x} & \mathbf{P}_{x\theta}\\\mathbf{P}_{x\theta}^T & \mathbf{P}_{\theta}\end{bmatrix},
\end{equation}
where $\mathbf{P}_{x}\in\Re^{m\times m}$ is the background error covariance computed from the forecast ensemble covariance of $x$, $\mathbf{P}_{x\theta}\in\Re^{m\times q}$ is the cross covariance between the model state $x$ and parameter $\theta$, and $\mathbf{P}_{\theta}\in\Re^{q\times q}$ is the background error covariance computed from the forecast ensemble covariance of $\theta$. The inference about the unobserved parameter and its uncertainty for the joint state-parameter estimation relies crucially on the cross covariance matrix $\mathbf{P}_{x\theta}$, which ``linearly regresses" the increment of the observed states to update the increment of the unobserved parameters. This can be easily seen by using the standard Kalman equation for the analyzed ensemble
\begin{equation}\label{eq:KFmean}
w^a=w^b+\mathbf{K}(y-\tilde{\mathbf{H}}w^b),
\end{equation}
where $w^b$ is the forecast state and the Kalman gain matrix $\mathbf{K}$ is given by
\begin{equation}\label{eq:KFgain}
\mathbf{K}:=\left[\begin{array}{c}\mathbf{K}_x\\\mathbf{K}_{\theta}\end{array}\right]=\mathbf{P}\tilde{\mathbf{H}}^T(\tilde{\mathbf{H}}^T\mathbf{P}\tilde{\mathbf{H}}^T+\mathbf{R})^{-1}.
\end{equation}
Substituting~\eqref{eq:Bk} into~\eqref{eq:KFgain}, we can rewrite ~\eqref{eq:KFmean} as
\begin{equation}\label{eq:KFmeansplit}
\begin{aligned}
x^a&=x^b+\mathbf{K}_{x}(y-\tilde{\mathbf{H}}x^b)\\
\theta^a&=\theta^b+\mathbf{K}_{\theta}(y-\tilde{\mathbf{H}}x^b),
\end{aligned}
\end{equation}

where the gain matrices for the model state $\mathbf{K}_k$ and the parameter $\mathbf{K}_\theta$ are given by
\begin{equation}\label{eq:KFmeansplit}
\begin{aligned}
\mathbf{K}_x&= \mathbf{P}_x\mathbf{H}^T(\mathbf{H}\mathbf{P}_x\mathbf{H}^T+\mathbf{R})^{-1}\\
\mathbf{K}_\theta&= \mathbf{P}_{x\theta}^T\mathbf{H}^T(\mathbf{H}\mathbf{P}_x\mathbf{H}^T+\mathbf{R})^{-1}.
\end{aligned}
\end{equation}
It is now clear that the gain matrix for the analyzed parameter ensemble depends on $\mathbf{P}_{x\theta}$. Notice that while the covariance matrix $\mathbf{P}_\theta$ has no effect on both gain matrices,$\mathbf{K}_x$ and $\mathbf{K}_\theta$, the covariance matrix $\mathbf{P}_x$ effects both. The equation for $\mathbf{K}_\theta$ in~\eqref{eq:KFmeansplit} also shows that the larger uncertainty in the forecast model state (i.e. the larger $\mathbf{P}_x$), the smaller rate change to the parameter increment for a given $\mathbf{P}_{x\theta}$. Therefore, in a chaotic system which typically causes a large ensemble spreading, the update of the model parameters can be expected to be slow. This may lead the filter divergence if the parameters are initially misspecified (e.g. the actual parameters are at the tail of the initial distribution of the parameters) since the parameters could be ``sticking" to their (incorrect) initial values for so long that the filter may end up repeatedly run forecasting with incorrect parameters, which lead most ensemble members to rapidly drift away from the observations. In some cases, some ensemble members may become dynamically unstable and the ensemble forecast becomes unbounded, in which case the filter ``blows up".

\section{Two-stages filtering}\label{sec:twostage}

Let consider again the combined state vector $w=[x,\theta]^T$. In Bayesian filtering framework, we aim to recursively evaluate the filtering distribution $p(w_k|y_{1:k})$, where $y_{1:k}=\{y_1,\ldots,y_k\}$. The particle filter (PF)~\cite{DGA2000} introduces an approximate solution to this problem without the assumptions of linearity or Gaussian uncertainties and they are not limited to estimating only the first two moments as in the Kalman-type methods. However, the PF is a computationally expensive method, which limits its applicability to a high-dimensional problem~\cite{SBBA2008}. Therefore, many modified PFs have been developed to reduce the overall computational load in comparison to the standard PF. The two-stages filtering proposed in this paper is motivated by an approach used in the Rao-Blackwellized particle filtering (RBPF)~\cite{RBPF99,DGA2000,DFMR2000} that runs PF on a part of state while updating the corresponding particles for the other part of state using the conditional KF. Suppose that the model state is evolved in a linear-Gaussian fashion, we may consider the following factorization for the joint state-parameter estimation:
\begin{equation}\label{eq:factor}
p(w_{1:k}|y_{1:k})=p(x_{1:k}|\theta_{1:k},y_{1:k})p(\theta_{1:k}|y_{1:k}).
\end{equation}
Although $p(x_{1:k}|\theta_{1:k},y_{1:k})$ is assumed to be Gaussian for a given set of parameters, $p(\theta_{1:k}|y_{1:k})$ is generally non-Gaussian. Running the standard PF for the combined state $w$ can be computationally expensive if the dimension of $w$ is large and it does not efficiently exploit the linear structure of the model state. The key idea of RBPF is that a PF method should be used on the parameter vector $\theta$, which is assumed to have a small dimension in this paper, while a KF method should be applied for the state vector $x$. To this end, the RBPF method approximates $p(\theta_{1:k}|y_{1:k})$ by weighted particles and rewrite Eq.\eqref{eq:factor} as
\begin{equation}\label{eq:RBPF}
p(w_{1:k}|y_{1:k})\approx\sum_{i=1}^{N}\omega_k^{(i)}p(x_{1:k}|\theta^{(i)}_{1:k},y_{1:k})\delta(\theta_{1:k}-\theta^{(i)}_{1:k}),
\end{equation}
where $\omega_k^{(i)}$ denotes the particle weight. Observe that $N$ KFs must be used to evaluate $p(x_{1:k}|\theta^{(i)}_{1:k},y_{1:k})$ in the above equation for each $i$. In general,$\theta_{1:k}$ can be sampled from any appropriate proposal density. For simplicity, we will sample $\theta_{1:k}^{(i)}$ from the transition density $p(\theta_k|\theta_{k-1})$, in which case we can use standard Bayesian analysis under some Markovian assumptions, see for exmaple,~\cite{AMGC2002}, to show that the particle weights can be recursively updated by
\begin{equation}\label{eq:PFweightupdate2}
\omega_{k}^{(i)}\propto p(y_k|y_{1:k-1},\theta_{k}^{(i)})\omega_{k-1}^{(i)}.
\end{equation}
Note that we do not naturally have a dynamic for the parameter, so we have to artificially design $p(\theta_k|\theta_{k-1})$. Some choices of the parameter dyanmics will be discussed later. The above predictive density of observation conditioned on the parameter paraticle serves as a likelihood function and it can be evaluated by
\begin{equation}\label{eq:PFlikelihood}
 p(y_k|y_{1:k-1},\theta_{k}^{(i)})\propto \mathcal{N}(H\hat{x}_k(\theta_k^{(i)}),HP^b(\theta_k^{(i)})_kH^T+R),
\end{equation}
where the mean estimate $\hat{x}_k(\theta_k^{(i)})$ and the background covariance $P^b(\theta_k^{(i)})$ are computed from the ensemble in the $i-$th KF.
It is clear that the computational cost per particle is generally more expensive than applying the standard PF on the combined state $w$. However, the RBPF can still be expected to improve the efficiency over the standard PF since fewer particles are required to achieve a given convergence~\cite{DGA2000,DFMR2000,SGN2005}. Also, the RBPF approaches have been reported to significantly reduce the variance of the particle weights in comparison to the standard PF~\cite{DFMR2000}. The two-stages filtering in this paper adopts the state partitioning apporach from RBPF but it reduces the overall computational load by ``approximating" the standard RBPF method as described below.
\par
Like the RBPF, the two-stages filtering uses PF for estimating the parameter vector$\theta$ and ENKF for the model state vector $x$. However, we replace $p(x_{1:k}|\theta^{(i)}_{1:k},y_{1:k})$ in Eq.~\eqref{eq:RBPF} by
\begin{equation}\label{eq:RBPFapprox}
p(x_{1:k}|\theta^{(i)}_{1:k},y_{1:k})\approx p(x_{1:k}|\hat{\theta}_{1:k},y_{1:k})
\end{equation}
This results in an ``interaction" of one PF and one single ENKF through a point parameter estimate from PF and the mean estimate of the state from ENKF; hence two-stages filtering. We also make an approximation
\begin{equation}\label{eq:PFlikelihoodapprox}
 p(y_k|y_{1:k-1},\theta_{k}^{(i)})\approx p(y_k|y_{1:k-1},\theta_{k}^{(i)},f(\hat{x}_{k-1}))\propto \mathcal{N}(H\hat{x}_k(\theta_k^{(i)}),R),
\end{equation}
where $\hat{x}_{k-1}$ is the state estimate from the ENKF in the $(k-1)-$th step. Without this approximation, one would have to run $N$ background updates from on the same analysis ensemble of the $(k-1)-$th step but for different parameter $\theta_k^{(i)}$. So, if the size of ensemble for ENKF is $M$, we have to compute $f_{k-1}^{k}(\cdot)$ in Eq.~\eqref{eq:model} for $MN$ times. The approximation in Eq.~\ref{eq:PFlikelihoodapprox}, however, reduces this computation to only $n$ times. Of course, there will be a loss in performance with these approximations. If $P(\theta_k|y_{1:k})$ is multi-modal, passing only the mean of this distribution to one single ENKF may result in filter divergence since the background ensemble in ENKF may diverge from a high probability region and likewise for passing only the mean of the state to the PF step. Therefore, we restrict our numerical experiments in the subsequent sections to the cases where the flow maps do not produce a multi-modal forecast distribution.
\par
The algorithm for the two-stages filtering is now be summarized below.
\begin{itemize}
  \item \textbf{Initialization}
  \begin{itemize}
    \item Sample initial particle for parameter $\theta^{(i)}_0$, $i=1,\ldots,N$
    \item Choose initial distribution for the state, say $\mathcal{N}(\hat{x}_0,P_0)$
    \item Sample initial state ensemble members $x^{(j)}_0$, $j=1,\ldots,M$
  \end{itemize}

  For every assimilation cycle $k$, we perform the following

  \item \textbf{PF-stage}
  \begin{itemize}
    \item Artificially ``move" the parameter particles according to some artificial (stochastic/deterministic) dynamic, say $g(\theta,\eta)$, and update the predicted observation
    \begin{equation}
    \begin{aligned}
    \theta^{(i)}_k &= g(\theta^{(i)}_{k-1},\eta_k),\quad\eta_k\sim \mathcal{N}(0,V)\\
     y_k^{(i)} &= h(f_{k-1}^{k}(\hat{x}_{k-1};\theta_{k}^{(i)}))+\varepsilon_{k}{(i)}\\
    \end{aligned}
    \end{equation}

    \item Compute the unnormalized weights in Eq.~\eqref{eq:PFweightupdate2} but using the approximation in Eq.~\eqref{eq:PFlikelihoodapprox}
    \item Normalized the weight to obtain the weighted particle $\{w_k^{(i)},\theta_k^{(i)}\}$
    \item If necessary, resampling $\{w_k^{(i)},\theta_k^{(i)}\}$
    \item Compute a point estimate $\hat{\theta}_k$ from $\{w_k^{(i)},\theta_k^{(i)}\}$ (e.g. ensemble mean)
  \end{itemize}

  \item \textbf{ENKF-stage}
  \begin{itemize}

    \item Update background ensemble to obtain the predicted observation
    \begin{equation}
    \begin{aligned}
        x_k^{(j)} &= f_{k-1}^{k}(x_{k-1}^{a,(j)};\hat{\theta}_{k})\\
        y_k^{(j)} &= h(x_{k}^{(j)})+\varepsilon_{k}^{(j)}\\
    \end{aligned}
    \end{equation}
    \item Using ENKF to obtain to find the analyzed distribution
     $\mathcal{N}(\bar{x}^a_k|\hat{\theta}_k, P^a_k)$ and analyzed ensemble $x_{k}^{a,(j)}$
    \item Set $\hat{x}_k=\bar{x}^a_k$
  \end{itemize}
\end{itemize}
\par
The choice of the artificial dynamic $g(\theta,\eta)$ also plays a pivotal role in success of this method.
Since the true parameters are assumed to be constant in time, the so-called persistence model has been commonly used in several studies and it is given by
\begin{equation}\label{eq:persistence}
\theta_{k+1}=\theta_{k}.
\end{equation}
However, if the initial parameter are misspecified to begin with or the ensemble size is too small, we can only resample from the same set of poorly informative particles at every assimilation cycles in the PF stage. This will eventually make the model state ensemble drifting too far away from the observation.
To overcome this issue, the persistence model may be replaced by a random walk model
\begin{equation}\label{eq:persistence}
\theta_{k+1}=\theta_{k}+\eta_k,\qquad\eta_k\sim \mathcal{N}(0,\mathbf{W}_k)
\end{equation}
for some given covariance matrix $\mathbf{W}_k$. This model is aimed to generate a new set of parameter particles at every assimilation cycle.
However,the independent random movement of parameter particles will result in parameter posteriors that is far too diffused since the covariance will increase over time. This issue has been long recognized and a solution has been proposed by Liu and West~\cite{LiuWest}. In their work, a new artificial model for the parameters is given by
\begin{equation}\label{eq:persistence}
\theta_{k+1} = \alpha\theta_{k}+(1-\alpha)\overline{\theta}_{k}+\eta_{t}\qquad0<\alpha<1,\\
\end{equation}
where $\overline{\theta}_{k}$ is the ensemble mean of the parameter ensemble. Clearly, this model is designed to ``shrinks" a new set of particles toward the mean at the degree determined by $\alpha$. Therefore, the over-dispersive issue of parameter particles is suppressed. In a frame work of the smoothing kernel, the optimal value of $\alpha$ can be calculated at each assimilation cycle for a given ``target" variance, which is typically the variance of parameter ensemble before applying any artificial dynamic to it, but this is usually inconvenient in practice and a heuristic choice of $0.95<\alpha<0.99$ may be used instead, see ~\cite{LiuWest} for more details.

\section{Case study 1: Lorenz-96 with parameterized forcing}\label{sec:case1}

In this numerical experiment, we assume that the dynamic of the``true" state is governed by the Lorenz-96 model~\cite{Lorenz96}
\begin{equation}\label{eq:Lorenz96}
\frac{dx_i}{dt} = (x_{i+1}-x_{i-2})x_{i-1}-x_i+F
\end{equation}
where $i=1,\ldots,N_x$ with cyclic indices and $F$ is the forcing function. We choose $N_x=40$ and assume that $F$ is parameterized by
\begin{equation}\label{eq:parameter96}
F= f_0+\theta_1\sin(\frac{2\pi}{\theta_2}i),
\end{equation}
where $f_0=8$ and $\theta=[\theta_1,\theta_2]$ is unknown and has to be estimated. The ``perfect model" case is assumed in this experiment, hence the forecast model also use~\eqref{eq:Lorenz96} and~\ref{eq:parameter96}. Therefore, the uncertainty in the unknown $\theta$ is the only source of the model error. Note that this setup allows us to justify our estimation skill by comparing the parameter estimates with the true parameter $\theta^\ast$, which is chosen to be $\theta^\ast=[2,40]$. In the presence of other sources of model errors, however, it would be better to emphasize parameter estimates that result in the model outputs fitting with the observations as well as possible, not the error in parameter estimates.
\par
The model~\eqref{eq:Lorenz96} is numerically solved by the fourth-order Runge-Kutta method with a time step $\Delta t=0.05$. We initialize the model state ensemble by running a spin-up run for $30,000\delta t$ and use the simulation from the next $6,000$ time steps in the experiment. One single member from the ensemble is then used as the ``truth" and the observations are constructed by adding the Gaussian noise with zero mean and covariance $0.1\mathbf{I}$ to the odd-indexed state variables, hence 20 observations. The parameter particles, however, has no ``climatological information", so we initialize the parameter $\theta$ according to $\theta_1\sim \mathcal{N}(4,1)$ and $\theta_2\sim \mathcal{N}(20,10)$, where the initial parameters are clearly misspecified and the standard deviation of the parameter ensemble is chosen to be large enough that the true parameter is ensured to be within the support of the initial ensemble.
\par
In the first experiment, we set the time interval between observations, denoted by $\delta t$, to $\delta=\Delta t$. This short interval between the assimilation cycle leads to an approximately linear flow map as previously demonstrated in~\cite{SB07}. An ensemble consisting of 250 particles is used in the augmented state method and the two-stages filtering uses 200 particles to estimate the parameter by the PF and 50 ensemble members to estimate $x_i$ by the ENKF. In second case, we extended the assimilation time interval to $\delta t=10\Delta t$, which elevates the degree of the nonlinearity of the flow map.

\begin{figure}[htbp]
\centerline{\includegraphics[scale=0.5]{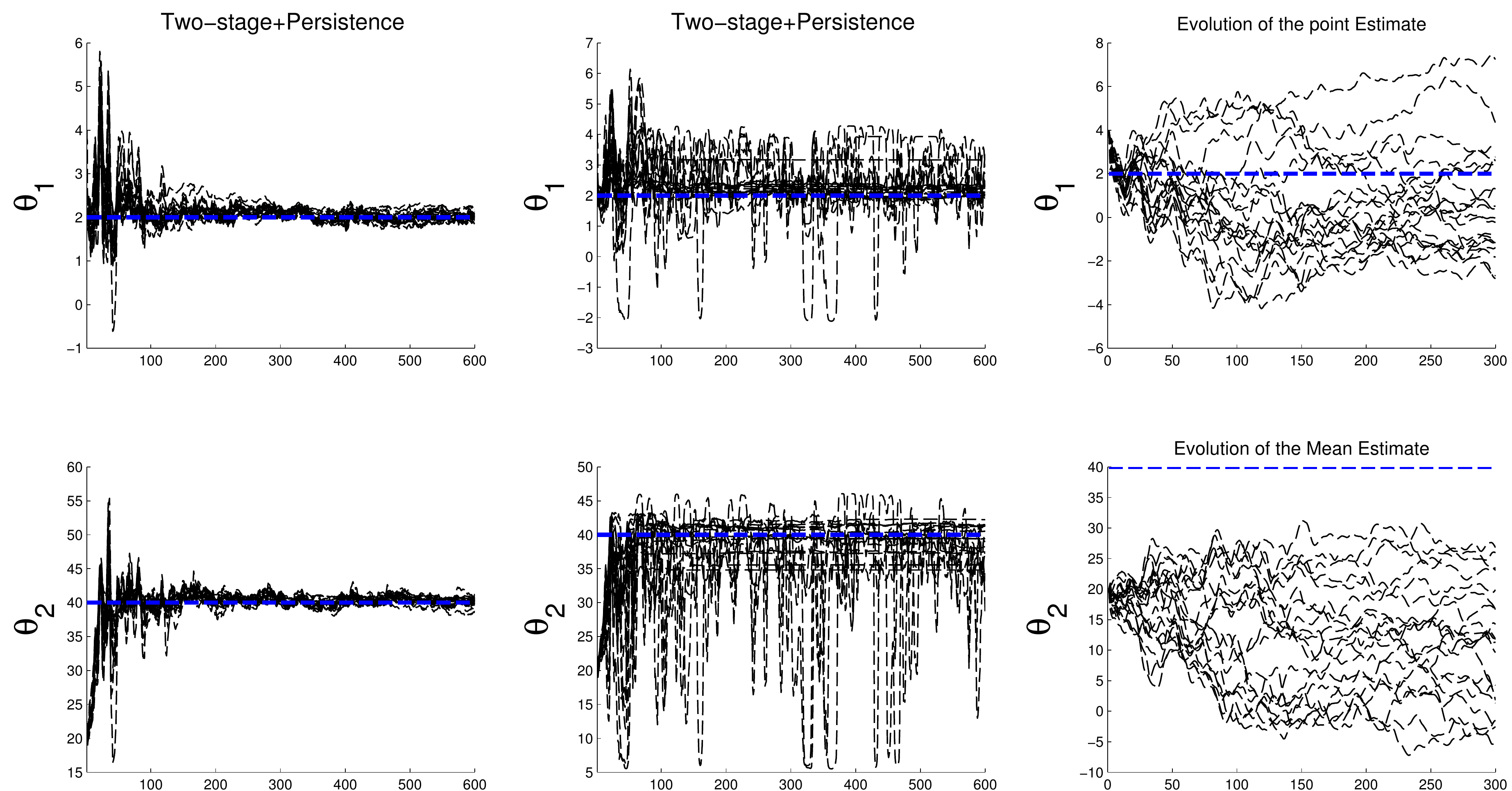}} \caption{The parameter posterior means from 20 different independent experiments for two-stagess filtering with 200 particles in the PF stage and 50 particles in the ENKF stage (Left) are compared with those obtained from the two-stages filtering with the same number of particles but using the persistence model instead of Liu-West model (Middle) as well as those from the ENKF (Right) with 250 ensemble members. The assimilation interval is $\delta=\Delta t$ and the true parameters, $\theta_1^\ast=4$ and $\theta_2^\ast=40$, are shown in the dash lines.}\label{fig:fluctuation_250pts_assim1}
\end{figure}

\begin{figure}[htbp]
\centerline{\includegraphics[scale=0.5]{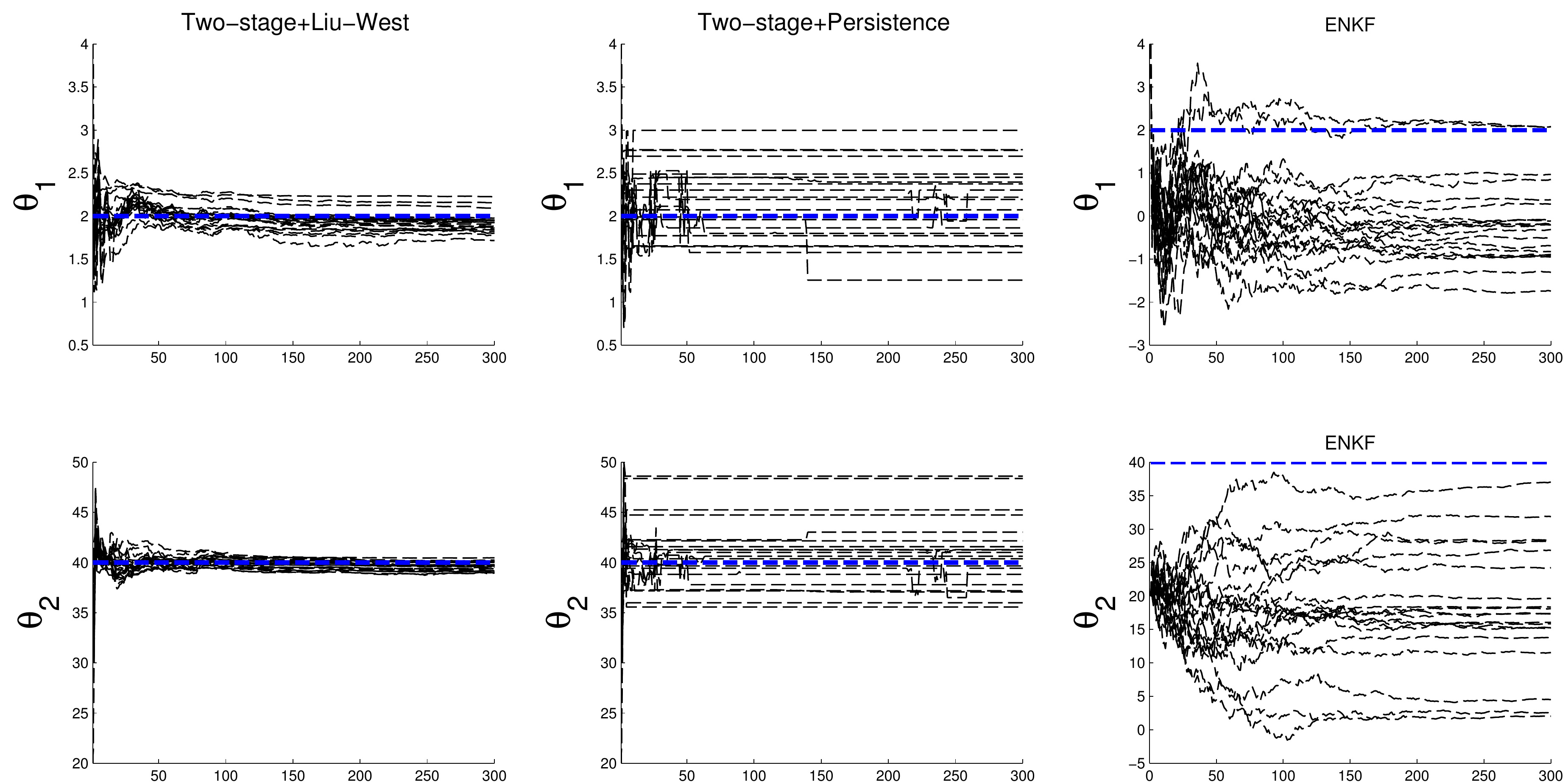}} \caption{Same as in Figure~\ref{fig:fluctuation_250pts_assim1} except $\delta=10\Delta t$.}\label{fig:fluctuation_250pts_assim10}
\end{figure}

\begin{figure}[htbp]
\centerline{\includegraphics[scale=0.5]{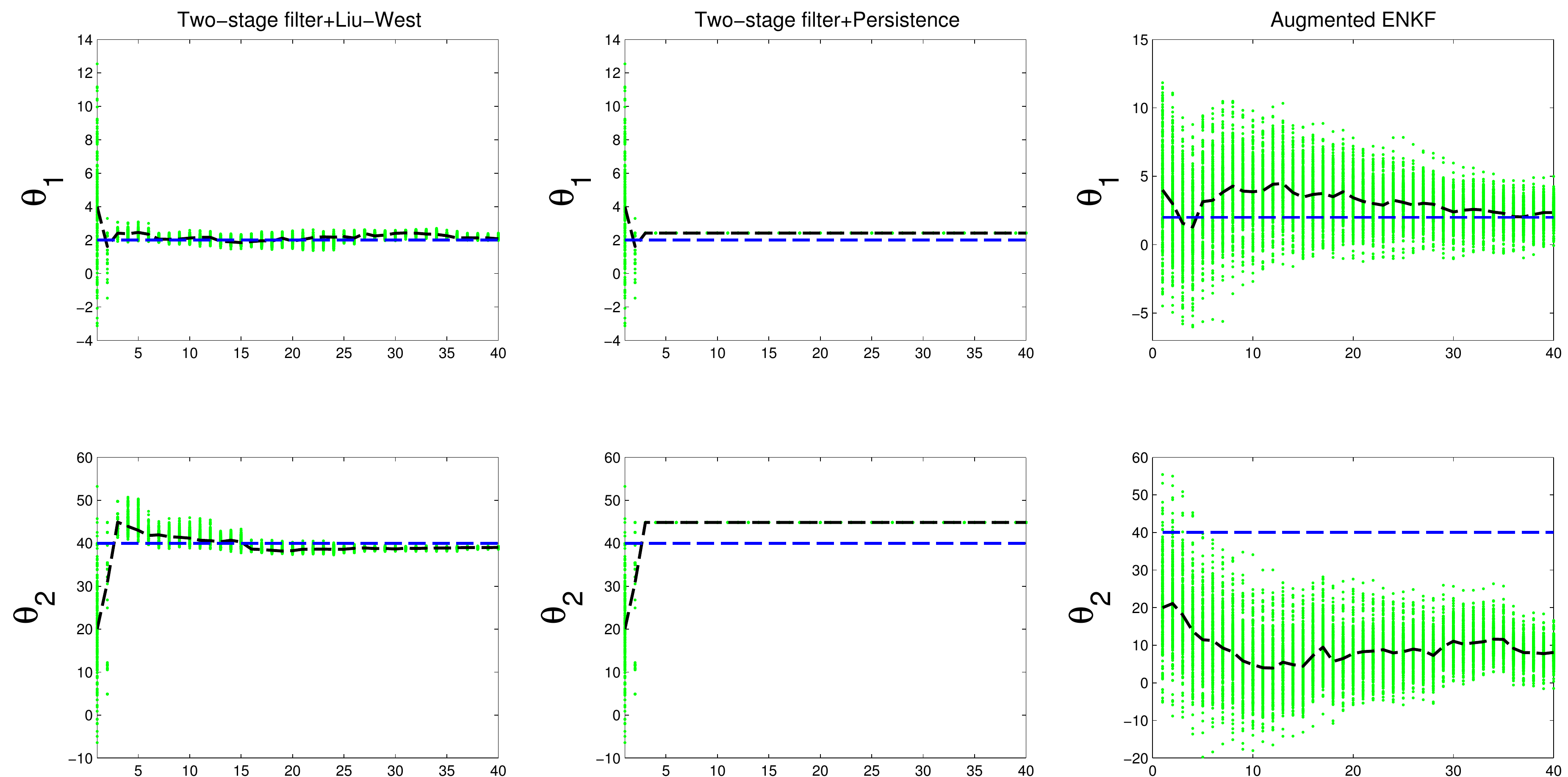}} \caption{The evolution of the ensemble (dots) and the ensemble mean (solid line) are shown as a function of the assimilation cycle. The dash line represent the true parameter. The ensemble size is 200 for the two-stage filtering methods and 300 for the augmented ENKF method. The assimilation time is $\delta t=10\Delta t$.}\label{fig:compareconvergenceL96}
\end{figure}

The mean estimates $\theta_1$ and $\theta_2$ for both two-stage filtering and augmented ENKF methods are compared in Figure~\ref{fig:fluctuation_250pts_assim1} and it is clear that the two-stages filter with the Liu-West model yields more accurate parameter estimates in all 20 different independent experiments, whose random initial ensembles are drawn independently. In the case of $\delta=10\Delta t$, similar results are obtained as shown in Figure~\ref{fig:fluctuation_250pts_assim10}. Figure~\ref{fig:compareconvergenceL96} compares the convergence of different methods for $\delta t=10\Delta t$ and the total number of 250 particles, of which 200 particles are used in the PF stage. It is clear that the two-stage filtering method with the persistent model is sensitive to the particle impoverishment, in which the particles become less diverse, and the convergence in the augmented ENKF method is relatively slow and the ensemble mean converges to an incorrect value of the parameter. As explained in Section~\ref{sec:Augmented}, the slow convergence of the augmented ENKF method can be explained by the large spreading of the ensemble of the state vector $x$ due to the large $\delta t$. In the case of the Liu-West model, the update of the ensemble mean is similar to that of the persistence model during the first few assimilation steps, after which  the ensemble in the persistence model collapses whereas the ensemble in the Liu-West model continues to explore a wider region in the parameter space.

\section{Case study 2: Fast-Slow Lorenz-96 system }\label{sec:case2}
We test our method in the case that the model includes the so-called ``closure parameter" that may arise from parameterizations of some unresolved physical processes. We consider the following fast-slow variant of Lorenz-96 model, where the slow variable $x_i$ is forced by the fast variable $y_{j,i}$.

\begin{equation}\label{eq:Lorenz96fastslow}
\begin{aligned}
\frac{dx_i}{dt} &= (x_{i+1}-x_{i-2})x_{i-1}-x_i+F-\frac{hc}{b}\sum_{j=1}^{N_y} y_{j,i}\\
\frac{dy_{j,i}}{dt} &= cb(y_{j-1,i}-y_{j+2,i})y_{j+1,i}-cy_{j,i}+\frac{hc}{b}x_iy_{N_y-1,i}
\end{aligned}
\end{equation}
where $i=1,\ldots,N_x$ and $j=1,\ldots,N_y$, both of which are cyclic. We use $F=8$, $N_x=16$, $N_y=8$, the coupling strength $h=1$, the time scale separation $c=10$ and the magnitude of the fast component $b=10$. It will be convenient to denote the fast-scale forcing by $g_i(x_i,t)=\frac{hc}{b}\sum_{j=1}^{N_y} y_{j,i}$. We use the fourth-order Runge-Kutta method to numerically integrate this fast-slow system with the integration time step $\Delta t=0.005$ to generate the ``truth" time-series for $x_i$ and $y_i$ and the observations are then constructed from $x_i$ for $i=1,3,\ldots,19$ by adding the realizations of the Gaussian distribution $\mathcal{N}(0,0.1)$.
\par
\begin{figure}[htbp]
\centerline{\includegraphics[scale=0.4]{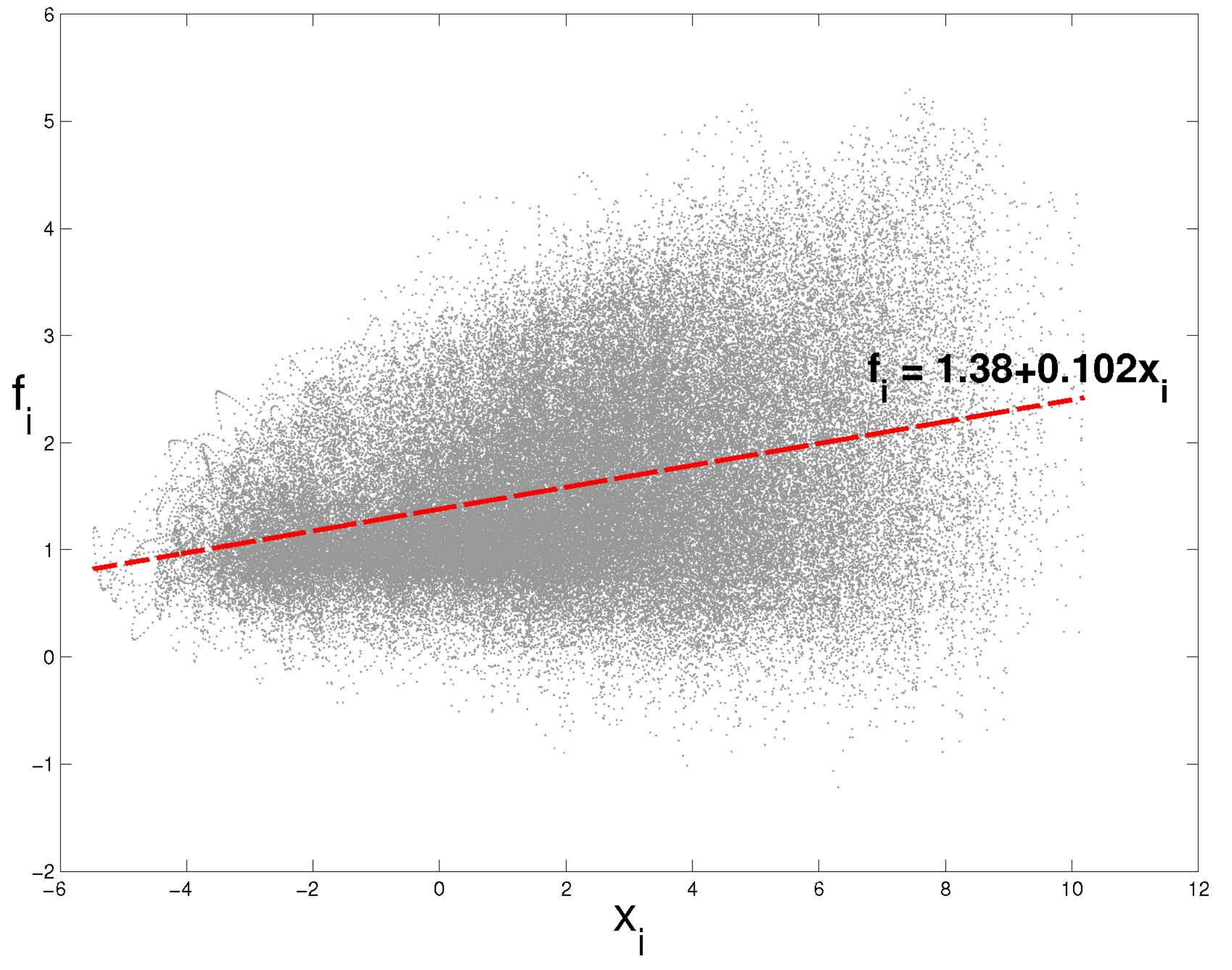}} \caption{A scatter plot of the true forcing $g_i$ and $x_i$ along with the straight line obtained by fitting $\theta_{1,2}$ to the pairs $g_i$ and $x_i$}\label{fig:scatter_fastslow}
\end{figure}

In the following experiments, we assume that only the physical process of slow variables is known, so we use a forecast model for the slow variables that takes into account the effect of the (unresolved) fast-scale variables only through a parametrization in term of the resolved variables $x_i$. In particular, the forecast model is given by

\begin{equation}\label{eq:Lorenz96slow}
\frac{dx_i}{dt} = (x_{i+1}-x_{i-2})x_{i-1}-x_i+F-(\theta_1(t)+\theta_2(t) x_i),
\end{equation}
where the polynomial with coefficients $\theta_1(t)$ and $\theta_2(t)$ represents an approximation of the unresolved forcing $g_i$. We plot the true forcing $g_i$ as a function of the true state variables $x_i$ using the truth and the result is shown as the scatter plot in Figure~\ref{fig:scatter_fastslow}. Clearly, in perspective of knowing only the slow process, there is an uncertainty in the fast-scale forcing $g_i$ for a given $x_i$, which is is higher for a large-scale $x_i$. Nevertheless, the trend of the data cloud in Figure~\ref{fig:scatter_fastslow} looks reasonably linear and the coefficients of the fitted line is found to to be $\theta_1^{\ast}=1.38$ and $\theta_2^{\ast}=0.102$. Therefore, we expect the mean of $\theta_1(t)$ and $\theta_2(t)$ to be close to $\theta_1^\ast$ and $\theta_2^{\ast}$, respectively.

\par
\begin{figure}[htbp]
\centerline{\includegraphics[scale=0.4]{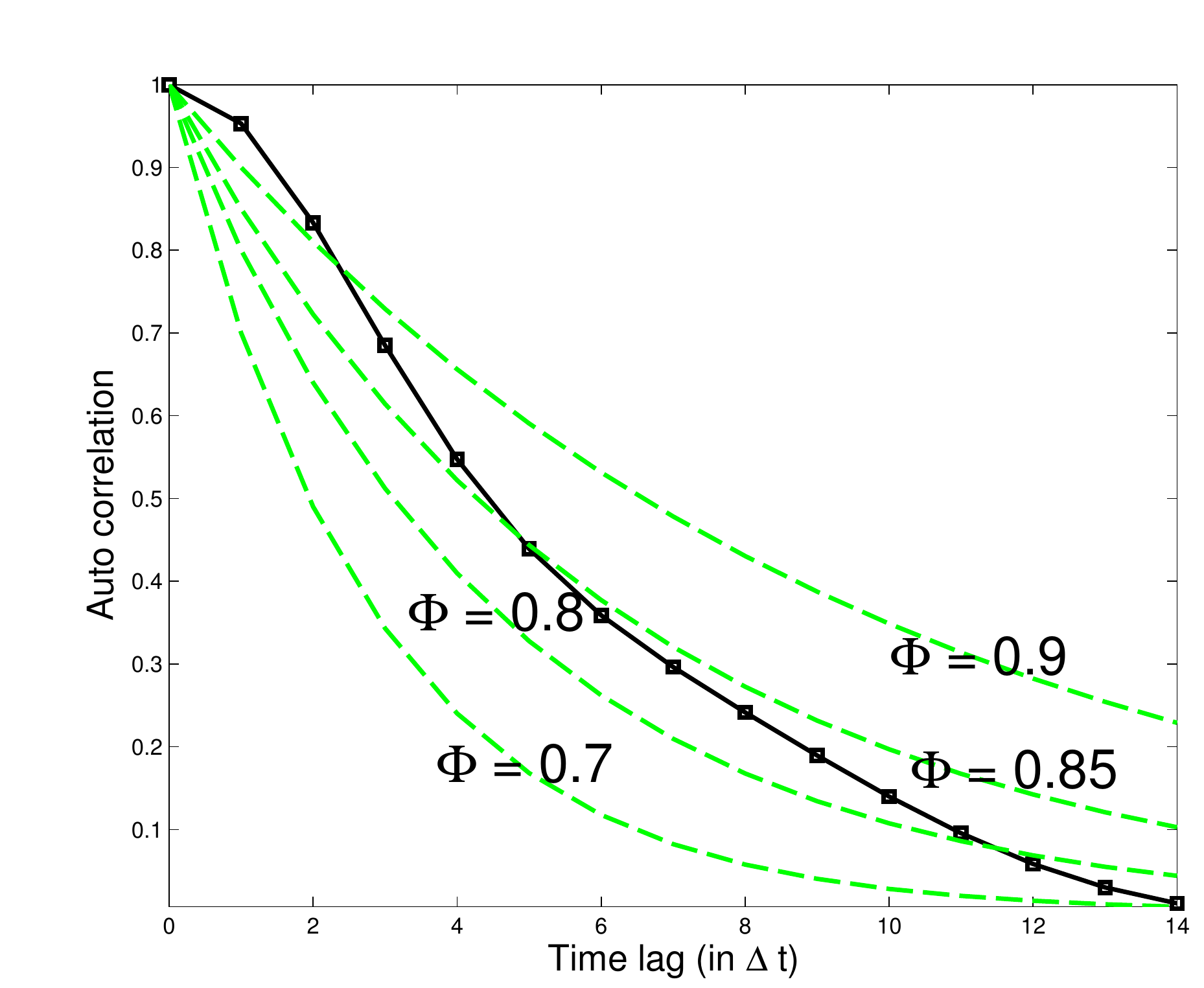}} \caption{Comparion between the acutral temporal autocorreltion of the true forcing and that implied by $e_i(t)$ in~\eqref{eq:AR1} for different values of $\phi$. }\label{fig:ACF}
\end{figure}

Since $\theta_1(t)$ and $\theta_2(t)$ are time-dependent, we will not directly estimate them. Instead, we will use the following models:
\begin{enumerate}
\item We consider~\eqref{eq:Lorenz96slow} as a random dynamical system where
\begin{equation}\label{eq:rnd1}
\begin{aligned}
&\theta_1(t)\sim\mathcal{N}(\mu_1,\sigma_1)\\
&\theta_2(t)\sim\mathcal{N}(\mu_2,\sigma_2)
\end{aligned}
\end{equation}
and try to estimate the parameters $\mu_{1,2}$ and $\sigma_{1,2}$. Note that our model for $\theta_{1,2}$ is imperfect in that it assumes $\theta_{1,2}$ to be uncorrelated and serially independent random forcing, and the variances $\sigma_{1,2}$ are independent of $x_i$, which contradicts with what was mentioned above. Nevertheless, we expect $\mu_{1}\approx\theta_1^{\ast}$ and $\mu_2\approx\theta_2^{\ast}$ but, as for $\sigma_{1,2}$, these values may or may not converge. For convenience, we will refer to this model as the ``random model" (RM).

\item We consider a stochastic parameterization
\begin{equation}\label{eq:stochasticpara}
\frac{dx_i}{dt} = (x_{i+1}-x_{i-2})x_{i-1}-x_i+F-(\theta_1+\theta_2 x_i)+e_i(t),
\end{equation}
where $e_i(t)$ is the stochastic forcing and represents uncertainty due to the deterministic parameterization.  Following a study of parametrizations in Lorenz's96 system by Wilks (2005), the deviation $e_i$ from the fitted line is given as an independent AR(1) process for each slow variable $x_i$:
\begin{equation}\label{eq:AR1}
e_i(t) = \phi e_i(t-\delta t)+\sigma_e(1-\phi^2)^{1/2}\eta_i(t),
\end{equation}
where $\eta_i(t)\sim\mathcal{N}(0,1)$. We will refer to the model in~\eqref{eq:AR1} as the ``stochastic model" (SM). Several parameter regimes were studied in Wilks but in this paper we will study a feasibility of using data assimilation to determine $\theta_1$, $\theta_2$,$\phi$, and $\sigma_e$. Again, we expect $\theta_1\approx\theta_1\ast$ and $\theta_2\approx\theta_2\ast$. However, the parameter$\sigma_e$ may or may not converge to a particular parameter since the model assumes $\sigma_e$ to be constant while the scatter plot in Figure~\ref{fig:scatter_fastslow} clearly shows that $\sigma_e$ depends on the slow variables $x_i$. As for the autoregressive parameter $\phi$, we also do not have a ``true" parameter to compare with since the temporal autocorrelation in the model decreases exponentially with time lag (in $\Delta t$) but the actual autocorrelation has a different trend as shown in Figure~\ref{fig:ACF}. In fact
, the study in Wilks (2005) demonstrated that a very wide range of values of $\phi$ and $\sigma_e$ yields similar results for the ensemble-mean RMSE.
\end{enumerate}

\subsection{Results}
\begin{figure}[htbp]
\centerline{\includegraphics[scale=0.5]{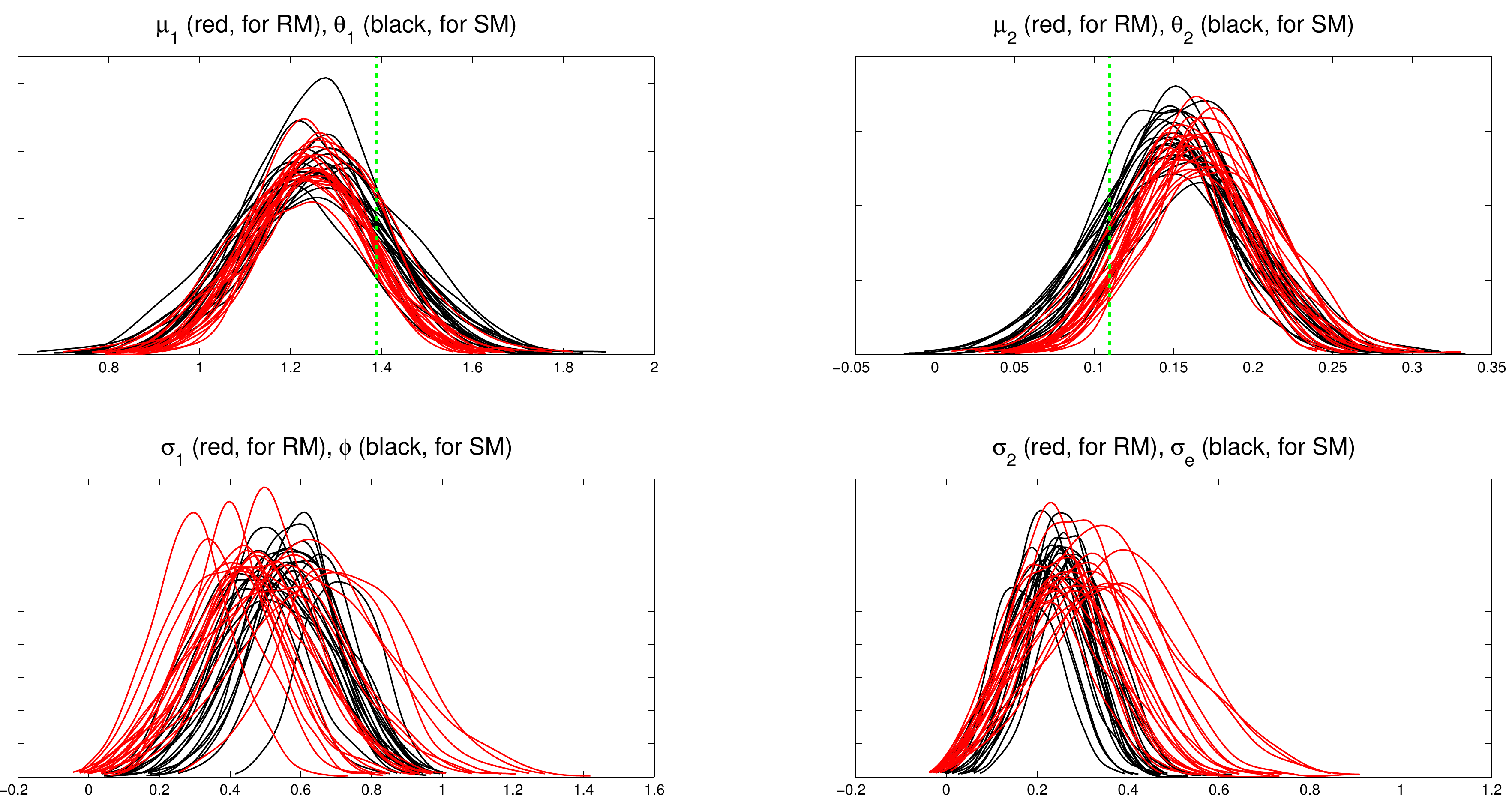}} \caption{The posterior distributions of the paramters after 400 assimilation cycles with $\delta t=50\Delta t$ from 20 independent experiments. The vertical lines in the top row show the values of $\theta_1^\ast$ (top-left) and $\theta_2^\ast$ (top-right).}\label{fig:L96fastslow_fitted}
\end{figure}

We run 20 independent experiments that starts with different initial ensembles drawn independently from the same prior distributions.
The parameters are initially drawn from the following prior distributions;$\mu_1\sim\mathcal{N}(3,2)$, $\mu_2\sim\mathcal{N}(-2,2)$, $\sigma_1\sim\mathcal{U}(0.1,0.9)$, $\sigma_2\sim\mathcal{U}(0.1,0.9)$, $\theta_1\sim\mathcal{N}(3,2)$, $\theta_2\sim\mathcal{N}(-2,2)$, $\phi\sim\mathcal{U}(0.1,0.9)$, and $\sigma_e\sim\mathcal{U}(0.1,0.9)$. We numerically solve~\eqref{eq:Lorenz96slow} with a numerical time step $\Delta t=0.005$ and set the assimilation cycle to $\delta=50\Delta t$. For the two-stage filtering, we use 200 particles in the PF stage and 50 ensemble members in the ENKF stage, hence using 250 ensemble members in the augmented ENKF for a fair comparison.
The marginal posterior distributions for the parameters of the RM and SM models after 400 assimilation cycles are compared in Figure~\ref{fig:L96fastslow_fitted} for the two-stage filtering. The distributions for $\theta_{1,2}$ for SM and $\mu_{1,2}$ for RM show similar characteristics in that their mean values are close to one another and their distributions all contain $\theta_1^\ast$ and $\theta_2^\ast$ in the supports. The means of the distribution of $\sigma_{1,2}$ spread out over the ranges of 0.3-0.7 for $\sigma_1$ and 0.2-0.4 for $\sigma_2$ whereas the means of $\phi$ and $\sigma_e$ vary in the ranges of 0.45-0.65 and 0.2-0.3, respectively. As for the augmented ENKF, the results are not shown here since it suffers the particle divergences, where most of particles diverge from the observation and the filter eventually ``blows-up", see again explanation in the end of Section\ref{sec:Augmented}

\begin{figure}[htbp]
\centerline{\includegraphics[scale=0.5]{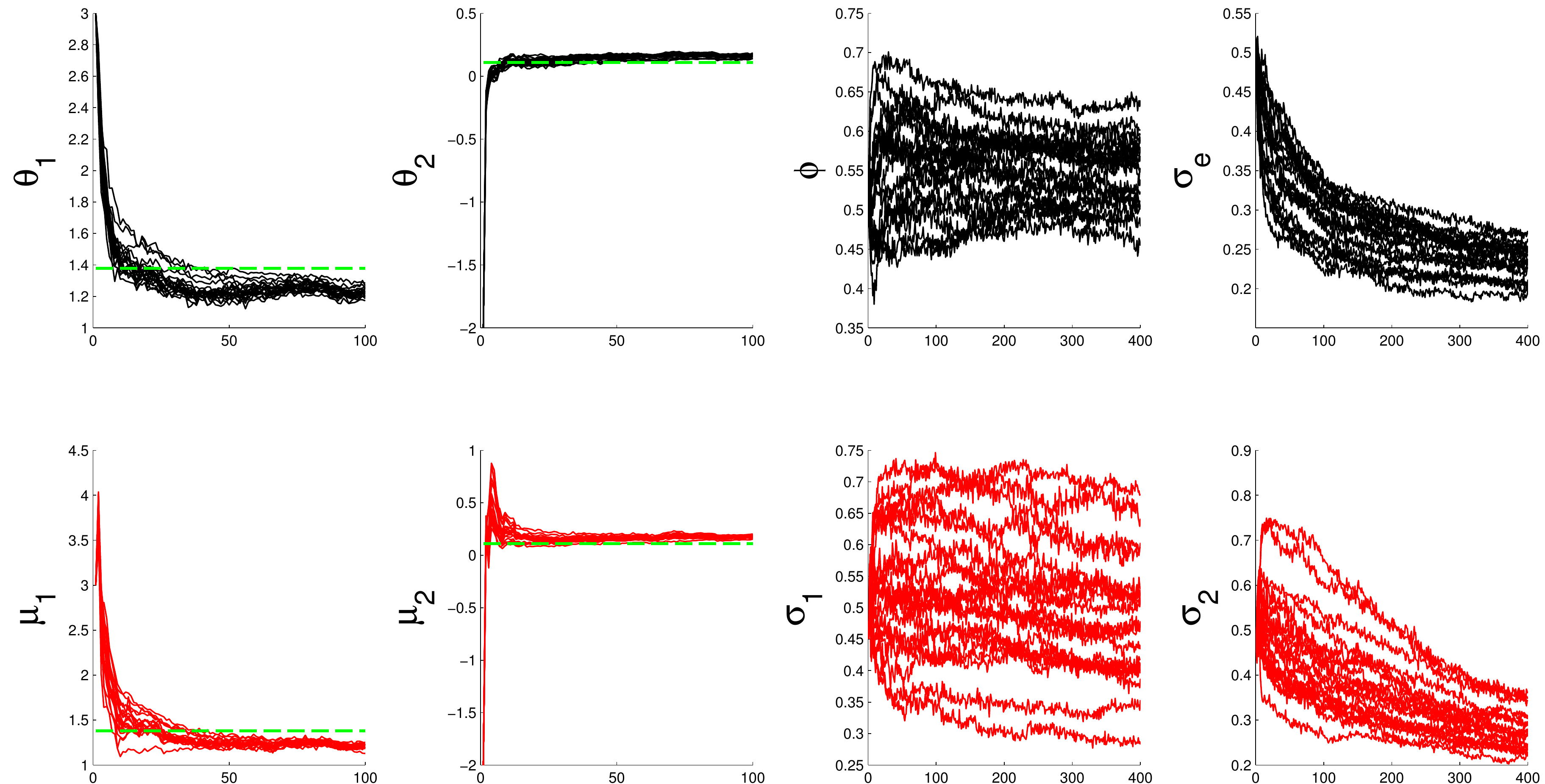}} \caption{The evolution of the mean eastimate for each parameters, see Figure~\ref{fig:L96fastslow_fitted} for the full distribution.}\label{fig:L96fastslow_meanonly}
\end{figure}
\par
\begin{figure}[htbp]
\centerline{\includegraphics[scale=0.5]{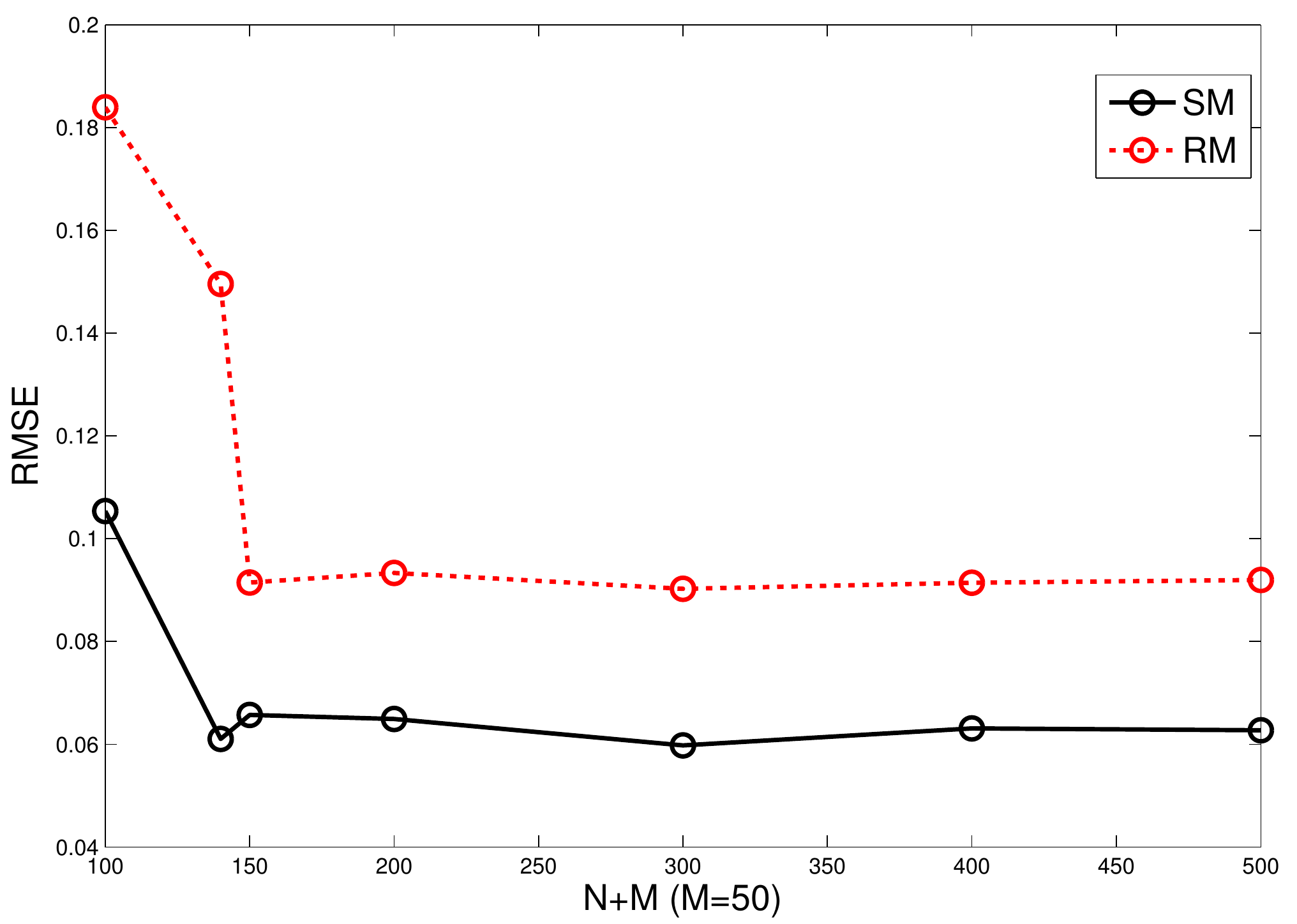}} \caption{RMSE for the two-stage filtering with the SM and RM parametrization schemes. The horizontal axis is the total number of particles, where $N$ is the number of particles in the PF stage and $M$ is the number of ensemble members in the ENKF stage.}\label{fig:compare_L96fastslow_err}
\end{figure}

In addition to comparing $\mu_{1,2}$ with $\theta^{\*}_{1,2}$, we measure the estimation skill by the error in the state estimates, which is given by
\begin{equation}\label{eq:rmse}
RMSE = \left\langle\frac{\sum_{i=1}^{N_x}\|x_i^{\text{true}}-\hat{x}_i\|_2^2}{\sum_{i=1}^{N_x}\|x_i^{\text{true}}\|_2^2}\right\rangle,
\end{equation}
\par
where $\langle\cdot\rangle$ means the average over the entire trajectory and all 20 different experimental runs. In Figure~\ref{fig:compare_L96fastslow_err}, the RMSEs in~\eqref{eq:rmse} obtained with both methods are compared for various total number of particles, for which the number of particles in the PF stage of the two-stage filtering method is fixed to $M=50$. Clearly, the RMSE is saturated for $N+M>200$.

\section{Summary and Discussions}
This paper proposed the two-stage filtering method for a joint state-parameter estimation based on a combination of the PF and ENKF methods. Specifically, the PF is used to estimate the uncertain parameter vector $\theta_k$ under an assumption that the initial state vector is known, using the mean of the analyzed state vector $\hat{x}_{k-1}$. The new parameter estimate is then updated based on the posterior parameter distribution approximated by the PF and used in the subsequent ENKF stage to update the state vector according to the Kalman update equations. Two numerical experiments are used to evaluate the ability of the two-stage filtering for the joint state-parameter estimation in comparison with the augmented ENKF method. Specifically, the first experiment uses the Lorenz 96 and assume that the forecast model and the parametrization is perfect and the parameter is constant. Partial observations (only half of the state variables is observed) and misspecified initial distributions of the parameters are used to test the proposed method in the ability to calibrating the incorrect parameters to the actual parameters. Our numerical results show that the two-stage filtering method yields more accurate parameter estimates that the augmented ENKF. The results also show the ineffectiveness of using the persistence model to artificially evolve the parameters. In particular, the use of the Liu-West model show a substantial improvement in the stability of the filter.
\par
The second experiment uses the fast-slow Lorenz 96 as the true model whereas the forecast model assumes the perfect physical law of the slow variable only but uses the first-order polynomial to parameterize the unresolved fast-scale variable. Two cases of the parametrization are used to test the proposed method. In the first case, the two coefficients of the polynomial are assumed to be (independent) realizations of the Gaussian process and we try to estimate the means and variances of the processes. The second case assume the two coefficients to be constant but adding the stochastic term into the forcing. This stochastic term is assumed to be a realization of an AR1 process, which is determined by the autocorrelation parameters and variance of the process. In both cases of the parameterizations, the mean of the actual parameters (i.e. the means of the Gaussian process in the first case and the constant coefficients in the second case) are properly estimated only by the two-stage filtering; the augmented ENKF blows up in the experiments. As for the other parameters, they converge for an individual run but not converge to the same parameter values when comparing 20 different independent experiments. Nevertheless, their uncertainties are reduced.
Further justifications of the accuracy of these parameter estimate will require an in-depth investigation of the range of the optimal parameter values for these two parameterization schemes.
\par
Since the applications of the two-stage filtering in this paper are restricted to the cases where the dimension of the parameter space is small and the parameters are spatially constant. In the case whose the parameter vector are spatially dependent and large dimensional, a localization technique may be needed to reduce the dimension of the original problem by using analyzing a local region with a smaller dimension. With this in mind, a localization for the PF must be developed and integrated into the two-stage filtering. In another situation where the flow map of the state vector produces a multi-model forecast distribution, the ENKF would be undoubtedly ineffective and we may have to replace the ENKF stage in the two-stage filtering by, for example, the PF. The above situations are beyond the scope of the present papaer and future works to deal with them will cer
tainly provide a better tool for the joint state-parameter estimation in the framework of the two-stage filtering than the one presented in this paper.

\section*{Acknowledgements}
This work was supported by the Office of Naval Research, grant number
N00014-11-1-0087 and NSF grant(s) numbered 1228265 and 0757527.

\bibliographystyle{unsrt}
\bibliography{TwoStageFiltering_Parameter_March2014}

 \end{document}